\documentclass[
showpacs,
floatfix,
aps,  
prl,
twocolumn,
superscriptaddress,
amssymb,
]{revtex4-1}

\usepackage{times}
\usepackage{wasysym}
\usepackage{amssymb}
\usepackage{latexsym}
\usepackage[dvips]{graphicx}
\usepackage{graphicx}
\usepackage{dcolumn}
\usepackage{amsfonts}
\usepackage{bm}
\usepackage{epsfig}

\newcommand{\be}{\begin{equation}}
\newcommand{\ee}{\end{equation}}
\newcommand{\bea}{\begin{eqnarray}}
\newcommand{\eea}{\end{eqnarray}}

\def\xt{X_T}

\def\m{m^\star}
\def\eac{\epsilon}

\def\oc{\omega_{\mbox{\scriptsize {c}}}}

\def\tq{\tau_{\rm q}}

\def\bo{\beta_{\omega}}

\def\mb{{m^\star_{\rm b}}}

\newcommand{\req}[1]{Eq.\,(\ref{#1})}

\newcommand{\rfig}[1]{Fig.\,\ref{#1}}

\newcommand{\rref}[1]{Ref.\,\onlinecite{#1}}

\def\ma{0.0662}
\def\na{1.26}

\def\mb{0.0644}
\def\nb{1.71}

\def\mc{0.0631}
\def\nc{3.16}

\def\ne{n_e}

\def\bo{B_\omega}

\def\ec{{\small$\Circle$}}
\def\fc{{\small$\CIRCLE$}}
\def\sq{{\small$\Square$}}

\begin{document}
\title{
Microwave-induced resistance oscillations in a back-gated GaAs quantum well
}
\author{X. Fu}
\affiliation{School of Physics and Astronomy, University of Minnesota, Minneapolis, Minnesota 55455, USA}
\author{Q. A. Ebner}
\affiliation{School of Physics and Astronomy, University of Minnesota, Minneapolis, Minnesota 55455, USA}
\author{Q.~Shi}
\affiliation{School of Physics and Astronomy, University of Minnesota, Minneapolis, Minnesota 55455, USA}
\author{M.~A.~Zudov}
\email[Corresponding author: ]{zudov@physics.umn.edu}
\affiliation{School of Physics and Astronomy, University of Minnesota, Minneapolis, Minnesota 55455, USA}
\author{Q.~Qian}
\affiliation{Department of Physics and Astronomy and Birck Nanotechnology Center, Purdue University, West Lafayette, Indiana 47907, USA}
\author{J.\,D. Watson}
\affiliation{Department of Physics and Astronomy and Birck Nanotechnology Center, Purdue University, West Lafayette, Indiana 47907, USA}
\author{M.\,J. Manfra}
\affiliation{Department of Physics and Astronomy and Birck Nanotechnology Center, Purdue University, West Lafayette, Indiana 47907, USA}
\affiliation{Station Q Purdue, Purdue University, West Lafayette, Indiana 47907, USA}
\affiliation{School of Materials Engineering and School of Electrical and Computer Engineering, Purdue University, West Lafayette, Indiana 47907, USA}
\received{March 21, 2017}

\begin{abstract}
We report on the effective mass measurements employing microwave-induced resistance oscillation in a tunable-density GaAs/AlGaAs quantum well.
Our main result is a clear observation of the effective mass increase with decreasing density, in general agreement with earlier studies which investigated the density dependence of the effective mass employing Shubnikov-de Haas oscillations. 
This finding provides further evidence that microwave-induced resistance oscillations are sensitive to electron-electron interactions and offer a convenient and accurate way to obtain the effective mass.
\end{abstract}
\received{21 March 2017}
\maketitle

It is well established that the effective electron mass $\m$ in GaAs/AlGaAs-based two-dimensional electron gas (2DEG) can deviate from the band mass of bulk GaAs, $m_b=0.067\,m_0$ ($m_0$ is the free electron mass).
One cause for this deviation is the non-parabolicity of the GaAs conduction band which leads to an enhancement of $\m$ with respect to $m_b$.
This enhancement becomes more pronounced at higher carrier densities and/or in narrower quantum wells. 
Another important aspect is electron-electron interactions which, depending on the carrier density $\ne$, can either increase or decrease $\m$ \citep{smith:1992,kwon:1994,coleridge:1996,tan:2005,zhang:2005b,asgari:2005,asgari:2006,drummond:2009,hatke:2013}.
Since cyclotron resonance is immune to interactions \citep{kohn:1961}, one usually resorts to $\m$ measurements using Shubnikov-de Haas oscillations (SdHO) to pick up these effects \citep{coleridge:1996,tan:2005}.

SdHO is a prime example of magneto-resistance oscillations which originate from Landau quantization when a 2DEG is subjected to a varying magnetic field $B$ and low temperature $T$. 
These oscillations owe to the commensurability between the Fermi energy and the cyclotron energy $\hbar\oc=\hbar eB/\m$.
Since these energies are both inversely proportional to $\m$, $\m$ cancels out and the SdHO frequency $B_{\rm SdHO} = \pi \hbar \ne /e$ can only be used to obtain the carrier density $\ne$. 
The information about $\m$ is contained in the SdHO amplitude which is proportional to $(\xt/\sinh \xt)\exp(-\pi/\oc\tq)$, where $\xt = 2\pi^2 k_BT/\hbar\oc \propto \m$, $k_B$ is the Boltzmann constant, and $\tq$ is the quantum lifetime.
Therefore, the only way to extract $\m$ from the SdHO measurements is through the examination of the decay of the SdHO amplitude with increasing temperature. 
Such approach, however, is very time consuming as it requires magnetoresistance measurements at several different temperatures followed by a careful analysis. 
Furthermore, the SdHO method suffers from a relatively low accuracy even if the data reduction procedure seems to work properly \citep{coleridge:1991,hayne:1992,coleridge:1996,hayne:1997}.
Therefore, it is very desirable to employ other experimental probes, which are free from the above drawbacks, to obtain $\m$.

One such probe is based on a phenomenon known as microwave-induced resistance oscillations (MIRO) which emerge in irradiated 2DEGs \citep{zudov:2001a,ye:2001}.
While MIRO also originate from Landau quantization, the role of the Fermi energy is now assumed by the energy of the incident photon $\hbar\omega$, where $\omega= 2\pi f$ is the microwave frequency.
As a result, the effective mass $\m$ can be obtained directly from the MIRO frequency,
\be
\bo = \frac {\m \omega} e\,,
\label{eq.bo}
\ee
 which does not contain any other unknown parameters and can be measured precisely in a single $B$-sweep.
In addition, it was recently shown \citep{hatke:2013} that $\m$ obtained using \req{eq.bo} differs from the value obtained from magneto-plasmon resonance \citep{hatke:2013}, indicating sensitivity of the MIRO mass to interaction effects.
Both of the above properties make MIRO an accurate, fast, and convenient option to investigate effective mass renormalization due to electron-electron interactions.

In this Rapid Communication we investigate the effect of the carrier density $\ne$ on the effective mass obtained from the MIRO frequency in a high-mobility GaAs/AlGaAs quantum well equipped with \emph{in situ} back gate.
At higher electron density ($\ne \approx 3.16 \times 10^{11}$ cm$^{-2}$), the analysis of the MIRO frequency revealed $\m < m_b$, in accord with \rref{hatke:2013}, which investigated MIRO in samples of similar density.
When the carrier density was lowered down to $\ne \approx 1.26 \times 10^{11}$ cm$^{-2}$, our MIRO data clearly showed an increase of $\m$.
While the increase of $\m$ is expected to occur with decreasing $\ne$, the detection of this increase previously required going to much lower densities \cite{tan:2005}, presumably, due to a considerably lower accuracy of the traditional SdHO approach.

Our 2DEG resides in a 30-nm GaAs/AlGaAs quantum well located about 200 nm below the sample surface.
The structure is doped in a 2 nm GaAs quantum well at a setback of 63 nm on a top side.
The \emph{in situ} gate consists of an $n^+$ GaAs layer situated 850 nm below the bottom of the quantum well \citep{watson:2015}.
The density of the 2DEG at zero gate bias is $\ne \approx 1.64 \times 10^{11}$ cm$^{-2}$ \citep{watson:2015}.
Ohmic contacts were fabricated at the corners and midsides of the lithographically-defined $1\times1$ mm$^2$ Van der Pauw mesa.
The low-temperature electron mobility varied from $\mu \approx 0.4\times 10^7$ to $\mu \approx 1.2\times 10^7$ cm$^2$/Vs over the density range studied.
Microwave radiation of $f = 34$ GHz, generated by a synthesized sweeper, was delivered to the sample immersed in liquid $^{3}$He via a rectangular (WR-28) stainless steel waveguide. 
The resistance $R$ was measured using a standard low-frequency (a few Hz) lock-in technique.

\begin{figure}[t]
\includegraphics{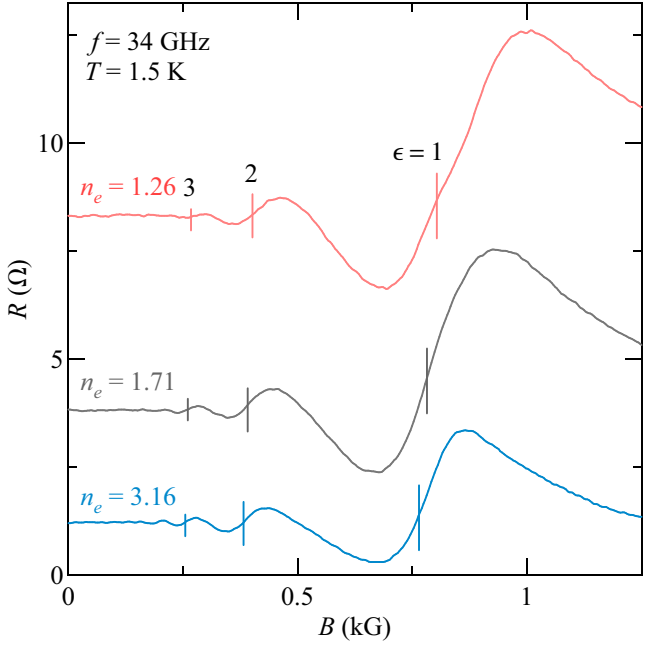}
\vspace{-0.1 in}
\caption{(Color online)
Magnetoresistance $R(B)$ measured at density $\ne \approx \na$ (top trace), $\nb$ (middle trace), and $\nc\times 10^{11}$ cm$^{-2}$ (bottom trace) at $T = 1.5$ K under irradiation by microwaves of $f = 34$ GHz.
Vertical line segments are drawn at $B = \bo/N$ for $N=1,2,3$, as marked.
}
\vspace{-0.1 in}
\label{fig1}
\end{figure}

Before presenting our experimental results, we recall that the radiation-induced correction to resistance which gives rise to MIRO can be described by \citep{dmitriev:2012,note:2}
\be
\delta R \propto - \lambda^2 \eac \sin 2\pi\eac \,, 
\label{eq.miro}
\ee
where $\eac \equiv \omega/\oc  \equiv \bo/B$ and $\lambda = \exp(-\eac/2f\tq)$ is the Dingle factor.
It then follows that the $N$-th order MIRO maximum ($+$) and minimum ($-$) can be described by  \citep{dmitriev:2012,note:2}
\be
\eac = \eac^{\pm}_N \equiv N + \delta_N^\pm \,,~~\delta_N^\pm \approx \mp 0.25\,,
\label{eq.ex}
\ee
and the $N$-th zero-response node, defined by $\delta R = 0$, by 
\be
\eac = N\,. 
\label{eq.nodes}
\ee 

While \req{eq.ex} is very simple, it should be used with caution. 
First, it follows from \req{eq.miro} which is valid only in the regime of overlapping Landau levels, i.e., when the amplitude of oscillations in the density of states (given by $\lambda \ll 1$) due to Landau quantization is small.
Second, it works best at low radiation intensities as high microwave power is known to reduce $|\delta_N^\pm|$ or even introduce additional oscillations \citep{hatke:2011e,shi:2017a}.
Finally, at sufficiently low values of $f\tq$, the exponential dependence of the Dingle factor can be strong enough to cause a significant shift of the oscillation extrema towards lower $\eac$ \citep{zudov:2014,shi:2014b,karcher:2016,shi:2017b}.
These considerations suggest that it is important to confirm that $|\delta_N^\pm| \approx 0.25$.
While none of the above limitations apply to \req{eq.nodes}, direct determination of the node positions from the experimental data is not possible.

In \rfig{fig1} we present magnetoresistance $R(B)$ for three different densities, $\ne \approx \na$ (top trace), $\nb$ (middle trace), and $\nc\times 10^{11}$ cm$^{-2}$ (bottom trace), measured at $T = 1.5$ K under irradiation by microwaves of $f = 34$ GHz.
It is evident that as the density is lowered, MIRO continuously shift to higher magnetic fields reflecting an increase of the effective mass.
The shift can also be discerned by comparing vertical line segments drawn at $B_N = \bo/N$ for $N = 1,2,3$, computed using \req{eq.bo} and $\m$ values obtained as discussed below.

\begin{figure}[t]
\includegraphics{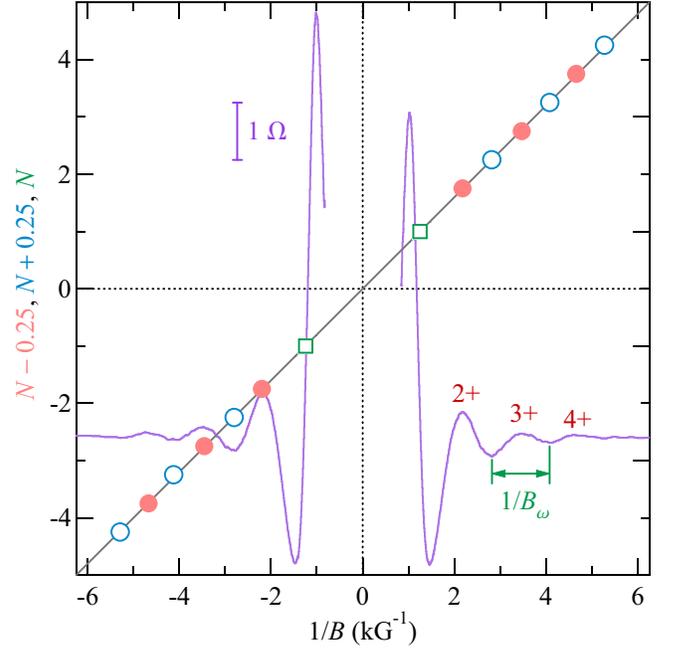}
\vspace{-0.1 in}
\caption{(Color online)
$N-0.25$ (\fc) and $N+0.25$ (\ec) as a function of $1/B$ at the MIRO maxima (cf. $2+,3+,4+$) and minima, respectively, obtained from $R$ (solid line) measured at $\ne = \na \times 10^{11}$ cm$^{-2}$, $T = 0.5$ K, and $f = 34$ GHz.
$N = \pm 1$ (\sq) vs $1/B = (1/B_1^+ + 1/B_1^-)/2$, see text.
Linear fit to $\bo/B$ generates MIRO frequency $\bo = 0.804$ kG, from which one obtains $\m  = \ma\,m_0$ using \req{eq.bo}. 
}
\vspace{-0.1 in}
\label{fig2}
\end{figure}
Since $\eac  = \bo/B \propto \m/B $, $\m$ can be readily obtained from the slope of $\eac_N^\pm$ vs $1/B$ evaluated at the MIRO extrema. 
This approach is illustrated in \rfig{fig2} showing $\eac_N^+$ (\fc) and $\eac_N^-$ (\ec) as a function of $1/B$ at the MIRO maxima (cf. $2+,3+,4+$) and minima, respectively, obtained from $R$ (solid line) measured at $\ne = \na \times 10^{11}$ cm$^{-2}$, $T = 0.5$ K, and $f = 34$ GHz.
One readily observes that the data points for both maxima and minima fall on the same straight line passing through the origin.
This observation is important as it confirms that the positions of the MIRO maxima are accurately described by \req{eq.ex}.
The linear fit (solid line) generates the MIRO frequency $\bo = 0.804$ kG, from which one obtains $\m  = \ma\,m_0$ using \req{eq.bo}.

While $|\delta_N^\pm| \approx 0.25$ is a good approximation for $N = 2,3,4$, the extrema near the cyclotron resonance are pushed towards the nodes at $\eac = \pm 1$ and are characterized by a considerably smaller $|\delta_1^\pm|$.
As a result, these extrema cannot be directly included in the analysis to obtain the mass.
However, since $|\delta_1^+| \approx |\delta_1^{-}|$, one can use the average position of these extrema, i.e., $1/B_1 = (1/B_1^+ + 1/B_1^-)/2$, to obtain data points at the node between them, $\eac = N = \pm 1$.
As shown in \rfig{fig2}, these points (\sq) are in excellent agreement with the rest of the data supporting the viability of the above approach.

\begin{figure}[t]
\includegraphics{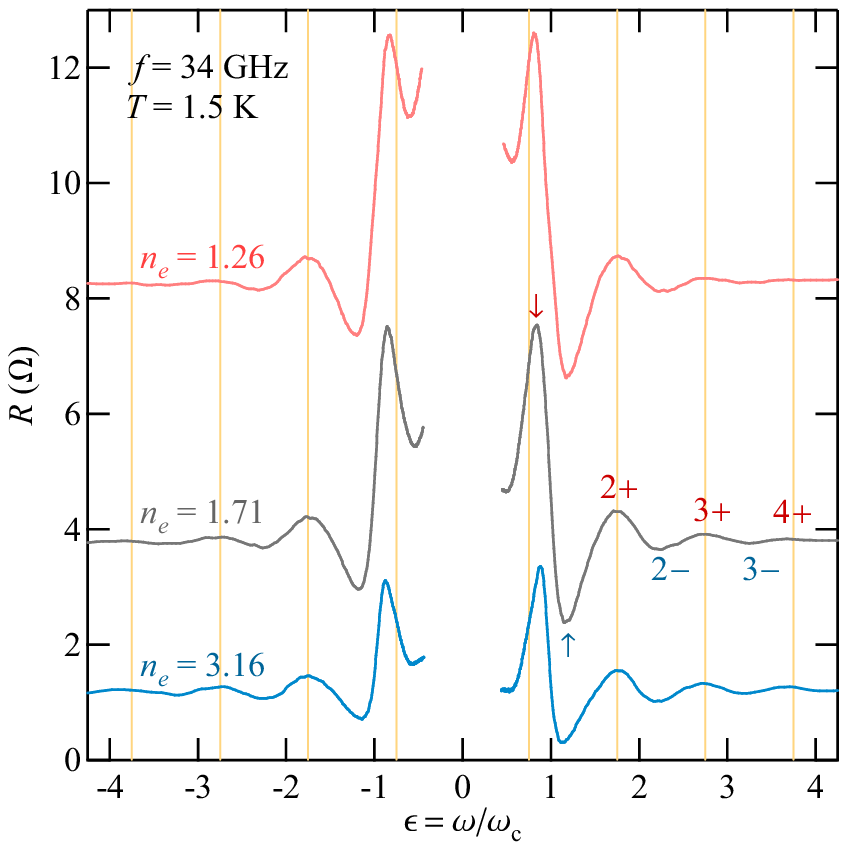}
\vspace{-0.1 in}
\caption{(Color online)
$R$ for $\ne \approx \na$ (top trace), $\nb$ (middle trace), and $\nc \times 10^{11}$ cm$^{-2}$ (bottom trace) measured at $T = 1.5$ K and $f = 34$ GHz as a function of $\eac =\omega/\oc$ computed using $\eac = \bo/B$ with $\m = \ma$, $\mb$, and $\mc \,m_0$, respectively.
MIRO maxima (minima) are marked by $N+$ ($N-$) for $N=2,3,4$ and by $\downarrow$ ($\uparrow$) for $N = 1$.
Vertical lines are drawn at $\eac = \pm (N - 1/4)$ for $N = 1,2,3,4$.
}
\vspace{-0.1 in}
\label{fig3}
\end{figure}

Having obtained $\bo$, it is straightforward to compute $\eac$ which allows further validation of the data reduction procedure to obtain the effective mass.
In \rfig{fig3} we present $R$ as a function of $\eac= \bo/B$ computed using $\m = \ma, \mb$, and $\mc\,m_0$ for $\ne \approx \na$ (top trace), $\nb$ (middle trace), and $\nc \times 10^{11}$ cm$^{-2}$ (bottom trace), respectively, measured at $T = 1.5$ K and $f = 34$ GHz.
Vertical lines are drawn at $\eac = \pm (N - 1/4)$ for $N = 1,2,3,4$.
These lines pass through all MIRO maxima with $|N|\ge 2$ confirming that $|\delta_N^+| \approx 0.25$. 
The same conclusion can be drawn for the MIRO minima.

\begin{figure}[t]
\includegraphics{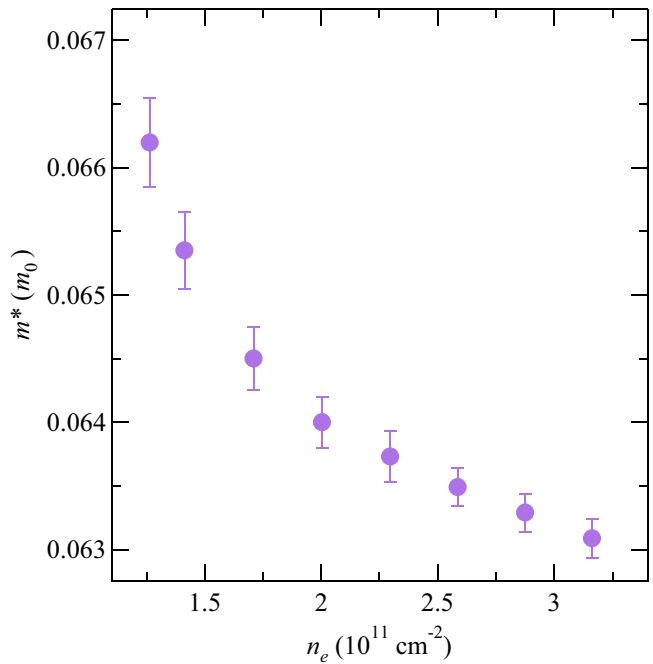}
\vspace{-0.1 in}
\caption{(Color online) Effective mass $\m$, in units of a free electron mass $m_0$, as a function of the carrier density $\ne$.
}
\vspace{-0.1 in}
\label{fig4}
\end{figure}

After repeating the effective mass extraction for other densities, we summarize our findings in \rfig{fig4} showing $\m$, in units of a free electron mass $m_0$, as a function of $\ne$.
We find that the effective mass increases \citep{note:3} from $\m \approx \mc$ to $\ma\,m_0$, as the density is lowered from $\ne \approx \nc$ to $\na\times 10^{11}$ cm$^{-2}$.
It is also evident that at lower $n_e$ the effective mass is changing at a faster rate.

It is interesting to compare our findings with an earlier study which investigated the density dependence of $\m$ obtained from SdHO in heterojunction-insulated gate field effect transistor (HIGFET) \citep{tan:2005}.
The findings of \rref{tan:2005} can be briefly summarized as follows.
At low densities, between $\ne \approx 1\times 10^{10}$ cm$^{-2}$ and $\ne \approx 1\times 10^{11}$ cm$^{-2}$, $\m$ showed a decrease from $\m \approx 0.085 - 0.1$ to $\m \approx 0.06 - 0.065$. 
However, further increase of density up to $\ne = 4 \times 10^{11}$ cm$^{-2}$ showed either little variation of the effective mass within the experimental uncertainty \citep{tan:2005} or a slight increase \citep{coleridge:1996} which could have originated from non-parabolicity \citep{note:3}. 
This is in contrast to our data which clearly show a noticeable decrease of $\m$ with increasing $\ne$ within this density range.

One possible reason for the above discrepancy is a much higher accuracy of our approach as compared to the SdHO analysis.
Indeed, the uncertainty of the mass obtained in \rref{tan:2005} is comparable to the mass change detected in our experiment.
However, it is also known that quantum confinement of a 2DEG under study sensitively affects mass renormalization due to electron-electron interactions \citep{zhang:2005b,asgari:2005,asgari:2006}.
More specifically, the finite thickness of the 2DEG softens the Coulomb interaction potential, resulting in a reduced mass value compared to the ideal 2D case \citep{zhang:2005b,asgari:2005,asgari:2006}. 
Furthermore, the dependence of the quantum confinement on the gate voltage is not universal but depends on the heterostructure design.
In contrast to HIGFET, the electron distribution in our quantum well becomes wider and more symmetric when a positive bias is applied to the back gate.
As a result, one should exercise caution when attempting quantitative comparison of our findings with that of \rref{tan:2005} or with existing calculations \citep{zhang:2005b,asgari:2005,asgari:2006}, both of which investigated a HIGFET realization of a 2DEG \citep{note:5}.

In summary, we investigated the effect of carrier density $\ne$ on the effective mass obtained from the MIRO frequency in a high-mobility modulation-doped GaAs/AlGaAs quantum well equipped with \emph{in situ} back gate over the density range from $\approx 1.2\times 10^{11}$ cm$^{-2}$ to $\ne\approx 3.2 \times 10^{11}$ cm$^{-2}$.
At the highest $\ne$, the analysis of the MIRO frequency revealed $\m \approx 0.063\,m_0$, considerably lower than the band mass value $m_b = 0.067\,m_0$, in qualitative agreement with \rref{hatke:2013} \citep{note:4}.
With decreasing density, the effective mass was found to increase exceeding $\m = 0.066\,m_0$ at the lowest density.
While the low-density increase of $\m$ has been previously established by SdHO measurements \cite{tan:2005}, it was detected only at much lower densities.
Taken together, our findings lend strong support that MIRO, like SdHO \citep{coleridge:1996,tan:2005}, are sensitive to electron-electron interactions but offer a much more convenient and accurate means to obtain $\m$. 
In addition, the MIRO approach can be directly applied to the effective mass renormalization studies in other systems, such as recently emerged high-quality Ge/SiGe and MgZnO/ZnO heterostructures.
Finally, our results are in general agreement with recent measurements of the MIRO mass in a series of individual samples covering a wider density range \citep{shchepetilnikov:2017}.

\begin{acknowledgements}
We thank I. Dmitriev for discussions.
The work at Minnesota (Purdue) was supported by the U.S. Department of Energy, Office of Science, Basic Energy Sciences, under Award \# ER 46640-SC0002567 (DE-SC0006671).
\end{acknowledgements}



\begin{thebibliography}{29}
\bibitem[{not({\natexlab{a}})}]{note:2}
\bibinfo{note}{For brevity, we limit theoretical expressions to the case of
  positive $B$, $N$, and $\eac$. For negative values, one should change sign of
  $\delta_N^\pm$.}
\end{thebibliography}
\end{document}